\documentclass[12 pt]{extarticle}
\usepackage{amssymb,amsmath,amsfonts,enumerate}
\usepackage{graphicx}
\usepackage{subcaption}

\usepackage[utf8]{inputenc}

\usepackage{epstopdf}
\usepackage{boxedminipage}
\usepackage{bm}
\usepackage{float}

\def\clock{{\count0=\time
           \divide\count0 60
           \ifnum\count0<10 0\fi\the\count0
           \multiply\count0 -60 \advance\count0 \time
           :\ifnum\count0<10 0\fi \the\count0
         }}
\newcommand{\timestamp}{{\small\vbox{\hbox{\tt\jobname.tex}
\hbox{\the\day/\the\month/\the\year, \clock}}}}

\newcommand{\beq}{\begin{equation}}
\newcommand{\eeq}{\end{equation}}
\newcommand{\ben}{\begin{displaymath}}
\newcommand{\een}{\end{displaymath}}
\newcommand{\beqa}{\begin{eqnarray}}
\newcommand{\eeqa}{\end{eqnarray}}
\newcommand{\bea}{\begin{eqnarray}}
\newcommand{\eea}{\end{eqnarray}}
\newcommand{\bean}{\begin{eqnarray*}}
\newcommand{\eean}{\end{eqnarray*}}
\newcommand{\ba}{\begin{array}}
\newcommand{\ea}{\end{array}}
\newcommand{\bi}{\begin{itemize}}
\newcommand{\ei}{\end{itemize}}

\numberwithin{equation}{section}

\begin{document}
\begin{titlepage}
\begin{flushright}
\end{flushright}
\vskip 2.cm
\begin{center}
{\bf\LARGE{Universal Basic Mass Density } }
\vskip 0.2cm

{\bf\LARGE{  inside Dark Matter Halos} }
\vskip 1.5cm

{\bf Nidal Haddad$^1$, Fateen Haddad$^2$}

\vskip 0.5cm
\medskip
\textit{ $^{1,2}$Department of Physics, Bethlehem University}\\
\textit{P.O.Box 9, Bethlehem, Palestine}\\
\textit{ $^{1,2}$Al-Quds Center for Physics}\\
\textit{Salah Al-Deen st. Addarmall Building, Jerusalem, Palestine}\\

\vskip .2 in
\texttt{haddad.nidal02@gmail.com, f.a.assaleh@gmail.com}

\end{center}

\vskip 0.3in

\baselineskip 16pt
\date{}

\begin{center} {\bf Abstract} \end{center} 

\vskip 0.2cm 

\noindent

In this work we propose a universal basic mass density and a universal basic metric inside dark matter halos in the framework of Einstein equations providing an analytical ground for learning about dark matter.
In a previous work the authors have proposed a simplified model for galaxies: a Schwarzschild black hole (that contains all the baryonic matter of the galaxy) immersed inside a dark matter halo. The solution solved the Einstein equations perturbatively and successfully gave the flat rotation curve and the baryonic Tully-Fisher relation, the two signatures of spiral galaxies. In this work we take the black hole mass (the baryonic mass) of the solution to be zero in order to focus our study on the dark matter halo exclusively. Among the results (1) is the prediction of a universal basic mass density of dark matter, $\rho=a_0/2\pi G r$, where $a_0\approx 2.8 \times 10^{-11} \ m/s^2$ is a universal constant whose value was deduced from observations, (2)  we show that the mass and velocity curve of the dark halo agree excellently with observational data at intermediate distances where the baryonic matter contribution is negligible, (3) we show that the constant $a_0$ is the origin of the constant surface density of dark matter and the origin of the scale-radius/scale-density scaling relation of the Navarro-Frenk-White profile (4) we show how the inclusion of baryonic matter in our model solves the core-cusp problem.
\end{titlepage} \vfill\eject

\setcounter{equation}{0}

\pagestyle{empty}
\small
\normalsize
\pagestyle{plain}
\setcounter{page}{1}

\newpage

\section{Introduction}

The study and analysis of observational data of rotational velocities of stars around galaxies and of velocity dispersions of galaxies in galaxy clusters have made it clear that most of the matter in these systems is hidden to us, in the sense that it does not send light. In general, this hidden matter does not interact with baryonic matter except through gravitation, being massive, and hence its effects are seen in the dynamics of luminous (visible) matter \cite{oort:1932,Zwicky:1933,Zwicky:1937,Rubin:1983}. This is the so-called dark matter (see, for example, the reviews \cite{Bertone:2016nfn,Garrett:2010hd,Arbey:2021gdg}). Put in another way, according to Newtonian gravity the luminous matter alone can not source the observed rotational velocity curves in galaxies and the velocity dispersions in galaxy clusters and therefore dark matter is needed in order to account for this discrepancy. On the cosmological scale  a similar kind of matter (dark energy) with even a larger percentage than dark matter is expected to fill the universe in order to account for the principal cosmological observations like the cosmological red-shift and the accelerating expansion of the universe, but this time according to Einstein's theory of general relativity. 

Finding the mass density of dark matter in galaxies and galaxy clusters was and is still one of the important goals in physics. One of the most successful and mostly accepted mass profiles of dark matter is the Navarro-Frenk-White profile (in short NFW) which was obtained by numerical simulations of dark matter according to the $\Lambda CDM$ model \cite{Navarro:1994hi,Navarro:1995iw}. It is worth emphasizing that the NFW numerical simulation did not include baryonic matter but only dark matter particles. The NFW density has proven to be successful in galaxies and galaxy systems fitting very well the observed rotation curves and velocity dispersions. It however does not succeed as much at short distances (in the cores of galaxies) where it stands at odds with observations; the NFW mass density predicts an $1/r$ density in the cores whereas observations show a constant density. This is known as the core-cusp problem -- see \cite{deBlok:2009sp,Flores:1994gz,Moore:1994yx,Tulin:2017ara,Weinberg:2013aya,Bullock:2017xww} and the references therein . One of the possible solutions to the core-cusp problem is that the inclusion of baryonic matter in the numerical simulations may change the situation in the core, that is, baryonic matter may back react and prevent the divergency in the density (see for example \cite{deBlok:2009sp,Navarro:1996bv,Pontzen:2011ty}).  

In contrast to the NFW profile, the impirical Burkert density profile \cite{Burkert:1995yz,Salucci:2018hqu} is another successful dark matter profile that is cored, that is,  it has an almost constant density in the core and does not suffer from the core-cusp problem, but, nevertheless, it is not very successful at large distances compared to the NFW profile. 

Scaling laws are important in observational astrophysics and cosmology since they may provide clues regarding the actual physical phenomena and their origins . By using the Burkert density function and observational data from a large sample of galaxies of different types and masses an interesting universal scaling relation has been found between the core radius ($r_c$) and the central density ($\rho_0$), the two fitting parameters of the Burkert profile \cite{Kormendy:2004se,Donato:2009ab,K:2020bmx}. This scaling law is equivalent to saying that the surface density of dark matter on the core surface has a universal value. The reason behind this constancy is still unknown however. A similar scaling relation has been found  as well between the two parameters of the NFW profile; in \cite{Boyarsky:2009rb,Boyarsky:2009af,DelPopolo:2012eb} the authors found that the product $\rho_s r_s$ is universal where a large and diverse sample of galaxies and galaxy clusters was studied with different types and masses.

In two pervious works \cite{Haddad:2020koo,Haddad:2020yhe} the authors of the current paper proposed a simplified model for galaxies based on General Relativity: a Schwarzschild black hole that contains all the baryonic mass of the galaxy immersed in a dark matter halo. The solution (1) solved the Einstein equations perturbatively, using the derivative expansion method, (2) successfully gave the flat rotation curve and the baryonic Tully-Fisher relation \cite{Tully:1977fu,McGaugh:2000sr}, the two signatures of spiral galaxies, (3)  showed a Keplerian curve before the onset of the flat rotation curve, (4) fulfilled the energy conditions of General Relativity  (strong, weak, null, and dominant \cite{Hawking:1973}) making it clear that the dark matter at hand is a proper physical field, (5) is consistent with the FLRW  spacetime in the sense that they can be put together consistently in a single framework (in a single gravitational metric), (6) provided an analytical ground for studying and predicting the relativistic behaviour of dark matter and its weak interaction with baryonic matter (7) showed that General Relativity is capable of solving the flat rotation curve problem without the need to be modified like suggested for example in conformal gravity \cite{Mannheim:2012qw} or mond \cite{Milgrom:2019cle} , and finally and most importantly for the current work (8) the solution predicted a universal constant with units of acceleration $a_0\approx 2.8\times 10^{-11}\ m/s^2$ whose value was deduced from observations.

In this work we take our model \cite{Haddad:2020koo,Haddad:2020yhe} and focus on the dark matter halo exclusively by taking the black hole mass to be zero and by so doing obtain a universal basic metric and a universal basic mass density for dark matter, namely, $\rho=a_0/2\pi G r$.  We start by calculating the total mass of dark matter inside the halo up to radius $r$ and the corresponding rotation curve and find that they agree excellently with the observational data analysis of \cite{Walker:2010ba,McGaugh:2006vv} where the authors collected observational data at intermediate distances (in galaxies) where the baryonic mass effect is negligible so as to focus on the dark matter exclusively. After that we show that the surface density of dark matter on the core radius $r_c$ is universal with almost the same central value predicted in the work \cite{Donato:2009ab} using observational data and the Burkert profile. Explicitly, we find that $\rho_0r_c=2a_0/\pi G =128\ M_\odot / pc^2$ where $\rho_0$ and $r_c$ are the central density and core radius respectively, the two fitting parameters of the Burkert profile. Similar results to the above were obtained as well by other authors \cite{Chavanis:2022vzi, deVega:2013woa}  using different models and ideas; in \cite{Chavanis:2022vzi} the author used the logotropic model of the universe \cite{Chavanis:2015paa} to predict similar results while in \cite{deVega:2013woa} the authors used the Thomas-Fermi approach for fermionic warm dark matter WDM  \cite{deVega:2017mcc,deVega:2013jfy}. 

Furthermore, we show that the basic density proposed in this work is the leading behaviour in the NFW profile and find a universal value for the product of the two fitting parameters of NFW: $\rho_s r_s=a_0/2\pi G=32\ M_\odot / pc^2$. This last scaling relation, which we call the scale-density/scale-radius scaling relation, was already found in \cite{Boyarsky:2009rb,Boyarsky:2009af,DelPopolo:2012eb} by analysing observational data and finding the best using the NFW profile. Moreover, we show how the inclusion of baryonic matter in our model stops the divergent $1/r$ behaviour in the basic density as one approaches the core and  by this we give further evidence for the expectation that the core-cusp problem can be solved by including the baryonic mass. Finally, we mention how the NFW profile (and other mass profiles in cosmology) can be seen as a cut-off to the basic density at large distances designed to make an end to the halo. We show also how the basic density connects smoothly with the Burkert profile (and other cored profiles) at the surface of the core. The last two sentences justify the choice of the name universal "basic" density. Thus, the aim of this paper is to give further evidence for the correctness and importance of the theoretical model proposed by the authors in \cite{Haddad:2020koo,Haddad:2020yhe} for galaxies and galaxy clusters as the model is shown to agree in an excellent way with observations.

\section {Model review: a black hole inside dark matter halo}
\label{review}
\begin{figure}[tbp]
\begin{center}
\includegraphics[bb=100 40 570 330 ,scale=0.7]{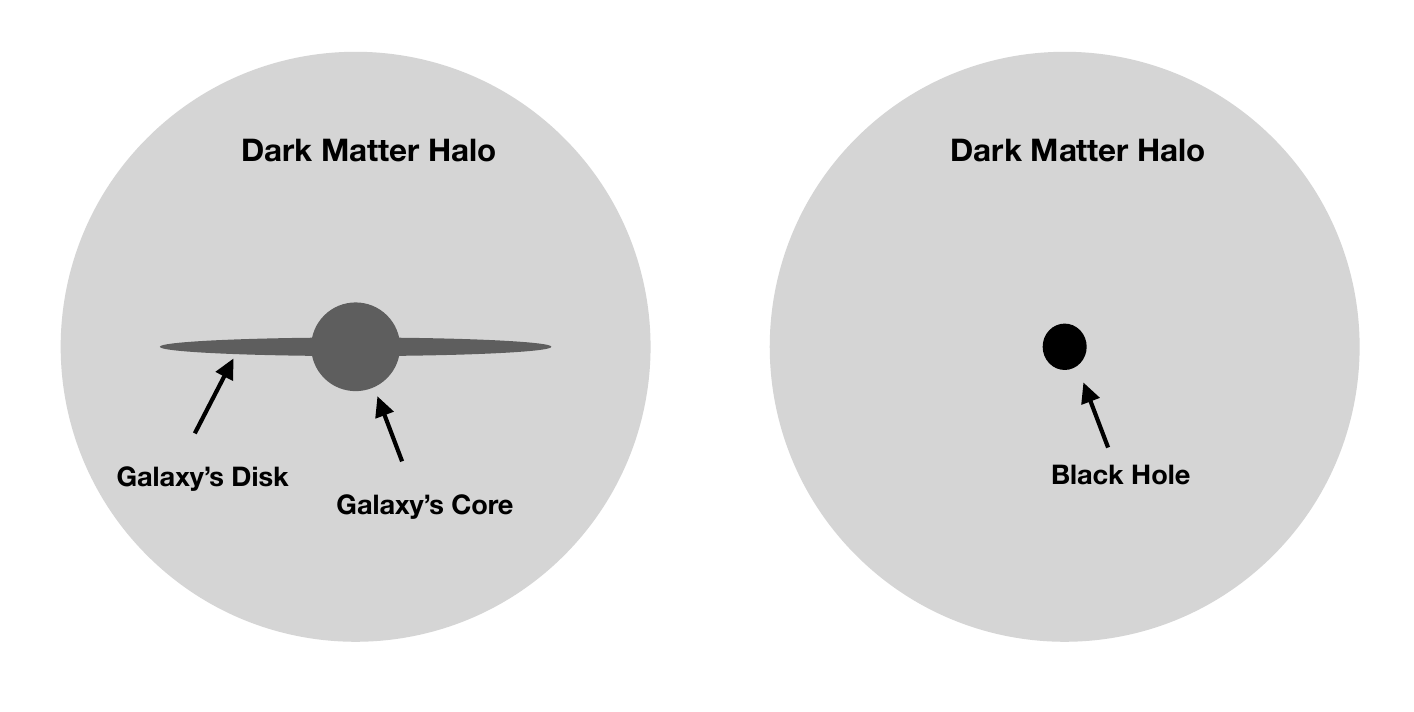}
\end{center}
\caption{\small Left: A spiral galaxy consisting of a core, a disk, and a dark matter halo. Most of the baryonic mass of the galaxy lies in its core. Right: Our simplified model for the galaxy consisting of a Schwarzschild black hole containing all the baryonic mass of the galaxy and the dark matter halo. Due to Gauss's theorem the two models give almost the same gravitational field at distances larger than the core radius; and the approximation gets better and better as we increase the distance.}\label{fig:halo}
\end{figure}

In this section we give a short review of the simplified model for galaxies proposed by the authors in \cite{Haddad:2020koo,Haddad:2020yhe} . A spiral galaxy (composed of a baryonic core, baryonic disk, and dark matter halo) is modelled as a Schwarzschild black hole that contains all the baryonic matter of the galaxy and is located in the centre of the dark matter halo (see Fig.\ref{fig:halo}). Since most of the baryonic mass of real galaxies is located in the core our model is very successful (very close to real galaxies) at distances larger than the core radius and it gets closer and closer to real galaxies as we increase the distance\footnote{Of course this is due to Gauss's law in gravity: the gravitational field outside any spherical  symmetric system is the same as the gravitational field of a point particle of the same mass located in the centre. }$^,$\footnote{In fact, our model can be applied to any type of galaxy, not only spiral ones, as long as most of the baryonic mass is enclosed inside the core. In the same token, our model can be applied equally well to galaxy clusters. }. 

Yet, even if the model is not close to real galaxies at short distances (in the cores) it can be considered as a theoretical simple system that can make us learn about dark matter and its properties more easily than more realistic models as they may be complicated and complex.

The spacetime metric of our galactic model is the following:
\beq\label{}
ds^2=-f(r)dt^2+h(r)dr^2+r^2d\Omega^2
\eeq 
where 
\beq\label{}
f(r)=1-\frac{2GM_b}{r}+2a_0(r-2GM_b)
\eeq 
\beq\label{}
h(r)=\left(1-\frac{2GM_b}{r}-2a_0(r-2GM_b)\right)^{-1}
\eeq 
where $M_b$ is the baryonic mass of the galaxy (the black hole mass), $d\Omega^2=d\theta^2+\sin^2\theta d\phi^2$, and the constant $a_0$ is universal for all galaxies, with units of acceleration and approximate value
 \beq
 a_0\approx2.8\times 10^{-11} \ m/s^2 
 \eeq
and it is the effect of  dark matter. The stress-energy tensor for this spacetime is:
\beq 
T_\mu^\nu=\textrm{diagonal}(-\rho,P_r,P_\perp,P_\perp)
\eeq
with 
\beq\label{density-P}
\rho=\frac{a_0}{2\pi G r}\left(1-\frac{GM_b}{r}\right)
 \qquad \quad P_r=-\frac{a_0M_b}{2\pi r^2} \qquad\quad P_\perp=\frac{a_0M_b}{4\pi r^2}
\eeq
which is anisotropic. This stress-energy tensor satisfies the four energy conditions of General Relativity (dominant, weak, null, and strong \cite{Hawking:1973}) outside the black hole horizon and thus the dark matter predicted by this model is a proper physical field with no unwanted features or illnesses.  

\begin{figure}[H]
\centering
\begin{minipage}[b]{0.45\linewidth}
\includegraphics[bb=0 0 250 380 ,scale=0.43]{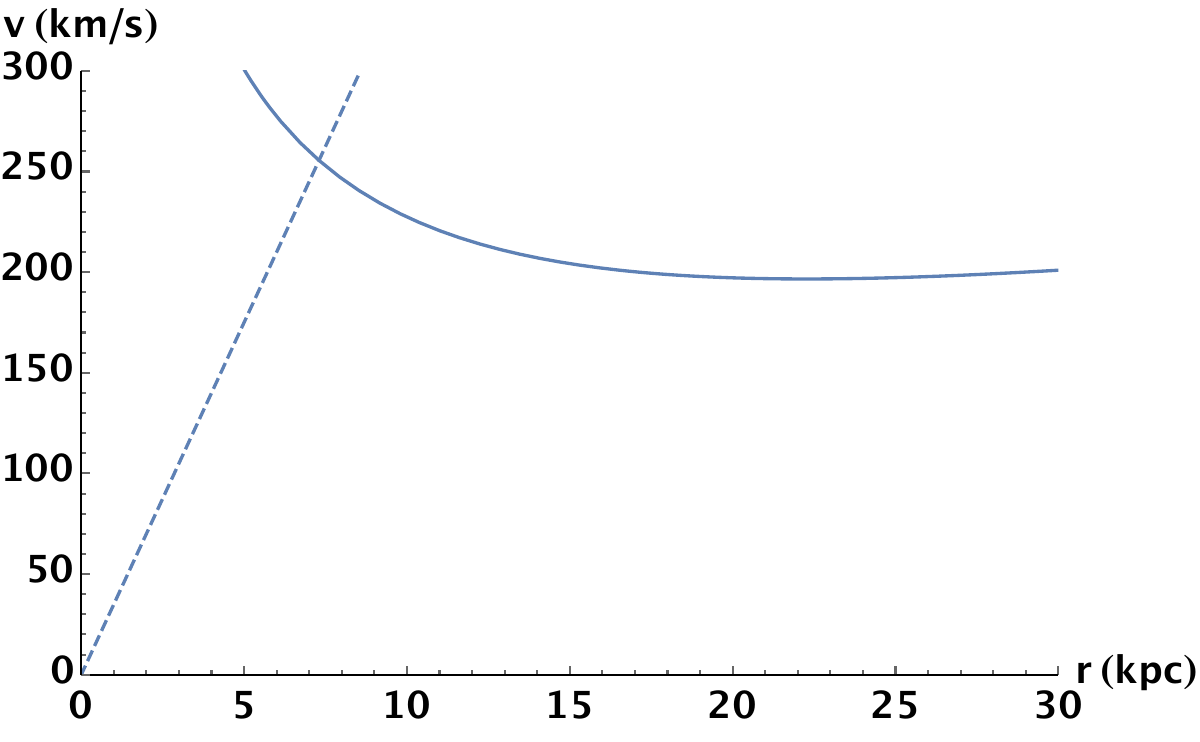}
\label{fig:minipage01}
\end{minipage}
\quad
\begin{minipage}[b]{0.45\linewidth}
\includegraphics[bb=0 0 250 380 ,scale=0.43]{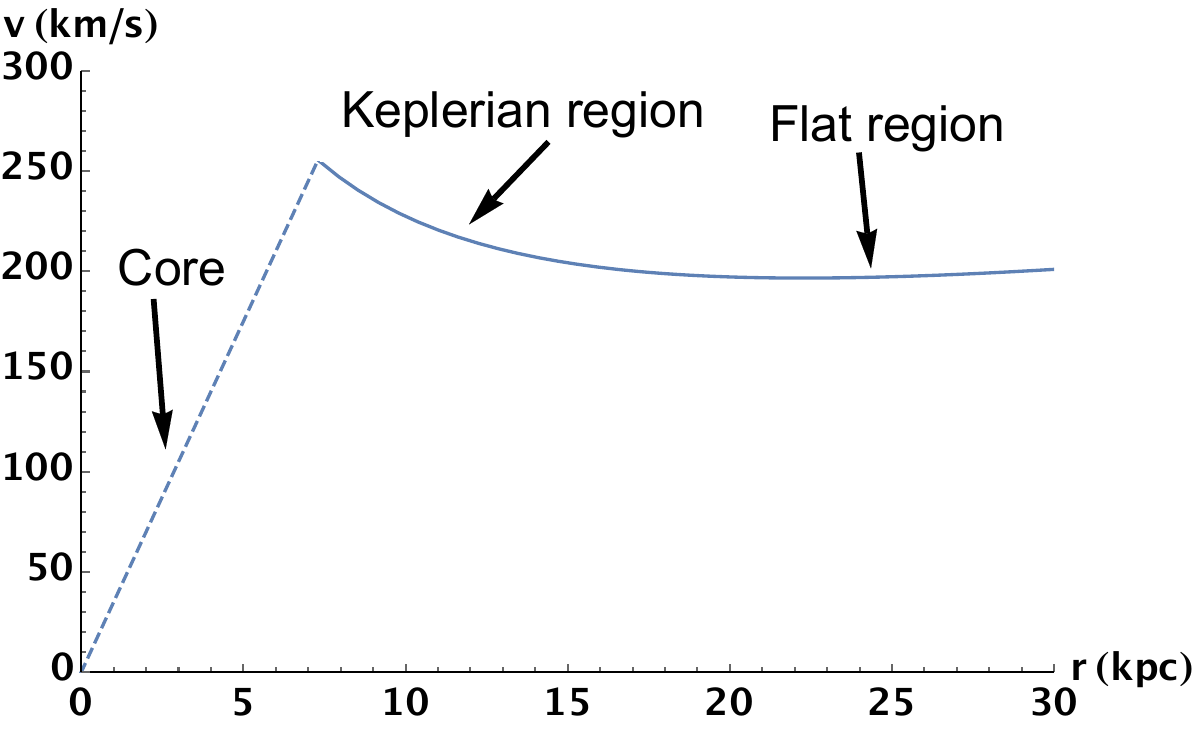}
\label{fig:minipage02}
\end{minipage}

\caption{\small Left: Our plot of the rotation curve of the milky way (solid line) with $M_b=1\times 10^{11} M_\odot$ and $a_0=2.8\times 10^{-11} \ m/s^2$ . The (dashed) straight line is a plot of the rotation curve of a rigid body, which is the accepted model inside the core which we add here for illustration only. Right: Since our plot is valid outside the core and the linear curve (straight line) is the accepted model inside, the two plots can be merged (in a rough way) by simply taking off the additional parts after their intersection.
 }

\label{fig:rot-curve}
\end{figure}

The rotation curve derived from this metric is given by the simple and compact equation:
 \beq\label{}
v=\sqrt{\frac{GM_b}{r}+a_0 r}
\eeq 
a plot of which is given in Fig.\ref{fig:rot-curve}  for the milky way. As can be seen the flat rotation curve appears along several kilo-parsecs; from almost 15\ kpc out to the edge of the galactic disk at 30\ kpc. Also the Keplerian rotation curve (namely, $v=\sqrt{GM_b/r}$) appears before the onset of the flat part.  

The model gives also the baryonic Tully-Fisher relation (see \cite{Tully:1977fu,McGaugh:2000sr}),

\beq\label{BTFR}
M_b=\frac{v^4}{4Ga_0} 
\eeq 
which states that the baryonic mass of any galaxy is proportional to the constant velocity to the power 4, with a universal constant of proportionality $1/4Ga_0$. Upon plugging in our value $a_0=2.8\times 10^{-11} \ m/s^2$ the proportionality factor of the baryonic Tully-Fisher relation  reads $1/4Ga_0=1.34\times 10^{20}\ kg\cdot\ s^4/m^4=67 M_{\odot}\ s^4/\ km^4$ which is close to the results of \cite{McGaugh:2000sr,McGaugh:2011ac,McGaugh:2005qe}.

\section {Universal basic density and metric for dark halos}
\label{}
After presenting the combined system (black hole plus dark halo) we can now focus our study on the dark halo alone by taking away the black hole from the picture by making $M_b=0$. Upon this the equations of the previous section  (Sec.\ref{review}) reduce to the following. The spacetime metric reduces to, 
 \beq\label{}
ds^2=-\left(1+2a_0r\right)dt^2+\frac{dr^2}{1-2a_0r}+r^2d\Omega^2
\eeq
which is the universal basic metric inside the dark halo. Here, $a_0\approx 2.8\times 10^{-11}\ m/s^2$ is the universal constant mentioned before. The term $a_0r$ as can be checked is very small for $r$'s of galactic scales and so we can expand the above metric and rewrite it as,
 \beq\label{met}
ds^2=\left(1+2a_0r\right)\left(-dt^2+dr^2\right)+r^2d\Omega^2
\eeq
Upon analysing the geodesics of this metric one sees that freely falling objects in this spacetime undergo a constant radial acceleration $a_0\approx 2.8\times 10^{-11}\ m/s^2$ toward the centre\footnote{This calculation must be done up to first order in derivatives \cite{Haddad:2020koo,Haddad:2020yhe}, or equivalently, up to $O(a_0)$. This can also be seen by taking the relevant equations in the references \cite{Haddad:2020koo,Haddad:2020yhe} and putting $M_b=0$.}.   

The energy-momentum tensor becomes,
\beq
T_\mu^\nu=\textrm{diagonal}(-\rho,0,0,0)
\eeq
where
\beq\label{basic density}
\rho=\frac{a_0}{2\pi G r}
\eeq
is the universal basic density of dark matter inside dark halos. This stress-energy tensor as can be seen describes a pressure-less gas with a mass density decreasing as $1/r$.
The Ricci scalar for this metric is,
\beq\label{}
R=\frac{4a_0}{ r}
\eeq
That this curvature scalar is divergent at the centre of the dark halo should not worry us (bearing in mind the cosmic censorship) since galaxies have baryonic super-massive black holes at their centres, which hide this singularity behind their horizons. 
Finally, the rotation curve reduces to a very simple equation,
 \beq\label{}
v=\sqrt{a_0 r}
\eeq 

\section {Mass, acceleration, and velocity inside dark halos} 
\label{}

According to the basic mass density of dark matter Eq.(\ref{basic density}) the total mass of dark matter up to radius $r$ is\footnote{The exact general relativistic integral must be $M(r)=\int_0^r\rho\times 4\pi r^2\sqrt{1+2a_0r}dr=\int_0^r \frac{a_0}{2\pi G r}\times 4\pi r^2\sqrt{1+2a_0r}dr$. However, we are interested in the answer up to order $O(a_0)$, and that is why the general relativistic factor  $\sqrt{1+2a_0r}$ can be neglected. Recall that the term $a_0r$ is very small in galaxies.}
  \beq\label{}
M(r)=\int_0^r\rho\times 4\pi r^2dr= \int_0^r \frac{a_0}{2\pi G r}\times 4\pi r^2dr
= \frac{a_0r^2}{G}\eeq
where one sees that the total mass is proportional to $r^2$. If we insert numbers ($a_0=2.8\times 10^{-11} \ m/s^2$) we obtain,
 \beq\label{}
M(r)=\frac{a_0r^2}{G}=202M_{\odot}\left(\frac{r}{pc}\right)^2
\eeq
where $\ pc$ stands for parsec, a result which agrees excellently with the  observational data analysed in reference \cite{Walker:2010ba} both in the $r^2$ behaviour and in the proportionality factor  (see also reference \cite{McGaugh:2006vv}). The authors in \cite{Walker:2010ba} analysed data from a large sample of spiral, dwarf spheroidal and low surface brightness galaxies at intermediate distances where the baryonic matter contribution is supposed to be negligible. The combined data set sampled dark matter halos over the range $r=0.2 \ kpc$ to $r=75\ kpc$. In addition, the authors in \cite{Walker:2010ba} did not adopt any halo model, so the comparison to their results consists a model-independent check on ours. Moreover, as shown in the previous section in this space-time there is a constant radial acceleration due to dark matter of magnitude $a_0\approx 2.8\times 10^{-11}\ m/s^2$ and a rotation curve $v=\sqrt{a_0 r}$; these two results agree also very nicely with the observational results obtained in \cite{Walker:2010ba,McGaugh:2006vv}.

The observational data \cite{Walker:2010ba,McGaugh:2006vv} have been taken at intermediate distances and not over the whole extent of the galaxies for two reasons. On the one hand, at short distances baryonic matter plays a significant role whereas the authors'  aim was to focus on the dark matter exclusively. On the second hand, at large distances the dark halo must naturally have an end somewhere and equivalently the basic density must have an end there too, and therefore a cut-off is needed; this cut-off is supposed to cut off the basic density together with the universal properties shown in this small section. As we are going to argue later, the NFW profile provides such a cut-off for the universal basic density Eq.(\ref{basic density}) (see Sec.\ref{cutoff}).

\section {Constant surface density of dark matter }
\label{burk}
One of the best-known mass profiles containing a core is the Burkert profile  \cite{Burkert:1995yz,Salucci:2018hqu},
 \beq\label{}
\rho_{Burk}=\frac{\rho_0r_c^3}{\left(r+r_c\right)\left(r^2+r_c^2\right)}
\eeq
where $\rho_0$ and $r_c$ are its two fitting parameters, called the central density and the core radius, respectively. It will be important for the following to notice that the density at the centre $r=0$ is $\rho_0$ whereas the density at the core radius $r=r_c$ is $\rho_0/4$. That is, the density at the core radius is one quarter of the central density.

Interestingly, there has been a growing observational evidence that these two constants (the central density $\rho_0$ and the core radius $r_c$) satisfy a universal scaling relation \cite{Kormendy:2004se,Donato:2009ab,K:2020bmx},

 \beq\label{}
\rho_0r_c=\text{constant}
\eeq
or in numbers (see reference \cite{Donato:2009ab}),
 \beq\label{dena}
\rho_0r_c=140^{+80}_{-30}\ M_\odot / pc^2
\eeq
where this product is proportional to the surface density of dark matter on the core surface. The authors in  \cite{Donato:2009ab} have found this scaling relation by best fitting the Burkert profile to a large number of galaxies of different types and masses\footnote{In fact, as stated in \cite{Donato:2009ab} the existence of a constant dark matter surface density was found to be independent of which cored profile was used.}. In more detail, the authors of \cite{Donato:2009ab} have obtained their results for galactic systems spanning over 14 magnitudes, belonging to different Hubble Types, and whose mass profiles have been determined by several independent methods.

In the following we show how we derive the above observational result from our model. Since the basic density Eq.(\ref{basic density}) approximates the real density very well outside the core and since the Burkert density approximates the real density very well inside the core one must have that the two densities are equal at the core radius $r_c$,
 \beq\label{}
\rho_{Basic}(r_c)=\rho_{Burk}(r_c)
\eeq
or explicitly,
 \beq\label{}
\frac{a_0}{2\pi G r_c}=\frac{\rho_0}{4}
\eeq
which gives,
 \beq\label{}
\rho_0r_c=\frac{2a_0}{\pi G }
\eeq
which, first of all, predicts a universal value for the product $\rho_0r_c$, and second, by plugging in numbers ($a_0=2.8\times 10^{-11}\ m/s^2$) we obtain
 \beq\label{128}
\rho_0r_c=\frac{2a_0}{\pi G }=128\ M_\odot / pc^2
\eeq
which very successfully fits in the observational result Eq.(\ref{dena}) of Ref.\cite{Donato:2009ab}. We would like to emphasize that this is done within the validity of our model which is supposed to be valid outside the cores of galaxies and therefore approaching the core $r=r_c$ from the outside, by decreasing the value of $r$, is still within the validity of our model. One can also look at this the other way around: the derivation of the observational result  Eq.(\ref{dena})  of Ref.\cite{Donato:2009ab} from our analysis shows that our model can indeed be trusted down to the core radius.

In the following we want to show that the result that the surface density of dark matter is universal is in fact not exclusive to the Burkert density. If we take any density which is constant (or almost constant) in the core, in accord with observations, then the full density can be written piecewise as,
\beq\label{}
\rho(r) =
 \left\{
     \begin{array}{lr}

                 \rho_0 &  r< r_c  \\
           a_0/2\pi Gr  & r\ge\ r_c \\

     \end{array}
   \right. 
\eeq
where we have taken the basic density to be the right model outside the core.  By asking that the two densities are equal at the core radius $r_c$ (i.e., the full density is continuous) one gets that,
 \beq\label{}
\frac{a_0}{2\pi G r_c}=\rho_0
\eeq 
and hence the scaling relation,

 \beq\label{32}
\rho_0r_c=\frac{a_0}{2\pi G }=32\ M_\odot / pc^2
\eeq
Note that the difference between the two values in Eq.(\ref{128}) and Eq.(\ref{32}) is due to the different profiles taken in the core; changing the profile changes the proportionality factor between the product $\rho_0r_c$ and the surface density. In any case, the surface density of dark matter turns out to be constant.

\section {Scale density - scale radius scaling relation  }
\label{NFW}

In this section we compare our basic density Eq.(\ref{basic density}) with the famous and mostly used Navarro-Frenk-White profile obtained from computer simulations \cite{Navarro:1994hi,Navarro:1995iw}, and applied to galaxies and galaxy clusters. The NFW density is given by 
 \beq\label{}
\rho_{NFW}=\frac{\rho_sr_s}{r\left(1+r/r_s\right)^2}
\eeq
where $\rho_s$ and $r_s$ are its two fitting parameters, where $\rho_s$ is called the scale density and $r_s$ the scale radius. 

If we expand the NFW profile in powers of $r/r_s$ we find that at leading order,

  \beq\label{}
\rho_{NFW}=\frac{\rho_sr_s}{r\left(1+r/r_s\right)^2}= \frac{\rho_sr_s}{r}+...            
\eeq
where one sees first that the $1/r$ behaviour is in common with ours. Second, by comparing to our profile Eq.(\ref{basic density})  we identify a relation between our universal constant $a_0$ and the two fitting parameters $\rho_s$ and $r_s$,
 \beq\label{}
\rho_sr_s=\frac{a_0}{2\pi G}=32\ M_\odot / pc^2
\eeq
which predicts a universal scaling law  -- see references \cite{Boyarsky:2009rb,Boyarsky:2009af,DelPopolo:2012eb} where the scaling law  between $\rho_s $ and $r_s$ was first found by best fitting the NFW profile to the data of a  large number of galaxies and galaxy systems of different types and masses.

 \section {Core-Cusp Problem} 
\label{}

The NFW profile is very successful on large scales (on galactic and galaxy cluster scales) but it suffers from some problems on small scales -- see for example the references \cite{deBlok:2009sp,Flores:1994gz,Moore:1994yx,Tulin:2017ara,Weinberg:2013aya,Bullock:2017xww}. One of these problems is that inside the cores of galaxies the NFW density grows sharply as we head for the centre in contrast to observations which show a constant density (i.e., a core). The contradiction between the divergent density (the $1/r$ behaviour) of NFW and the constant density observations is what is known as the core-cusp problem (see the references just mentioned). 

It is to be said that all the simulations that are based on the $\Lambda$CDM model suffers from the core-cusp problem, like the NFW, but it is to be said also that these simulations include only dark matter particles and do not include baryonic matter. Therefore one of the proposed solutions to the core-cusp problem is that the inclusion of baryonic matter may solve the problem and break the divergency \cite{deBlok:2009sp,Navarro:1996bv,Pontzen:2011ty} .

As for our profile Eq.(\ref{basic density}), it also does not have a core but as we have mentioned before our model by construction is supposed to approximate excellently real galaxies only outside the cores. To complete the picture we have to sue (match) consistently our solution with the appropriate solution inside the core (for instance with Burkert density) at the surface of the core -- see Sec.\ref{cutoff}. Nevertheless, as a toy model in the core,  we can take our full solution (baryonic black hole plus dark halo, reviewed in Sec.\ref{review}) and see what it has to say regarding the core-cusp problem.

\begin{figure}[tbp]
\begin{center}
\includegraphics[bb=100 20 400 300 ,scale=0.7]{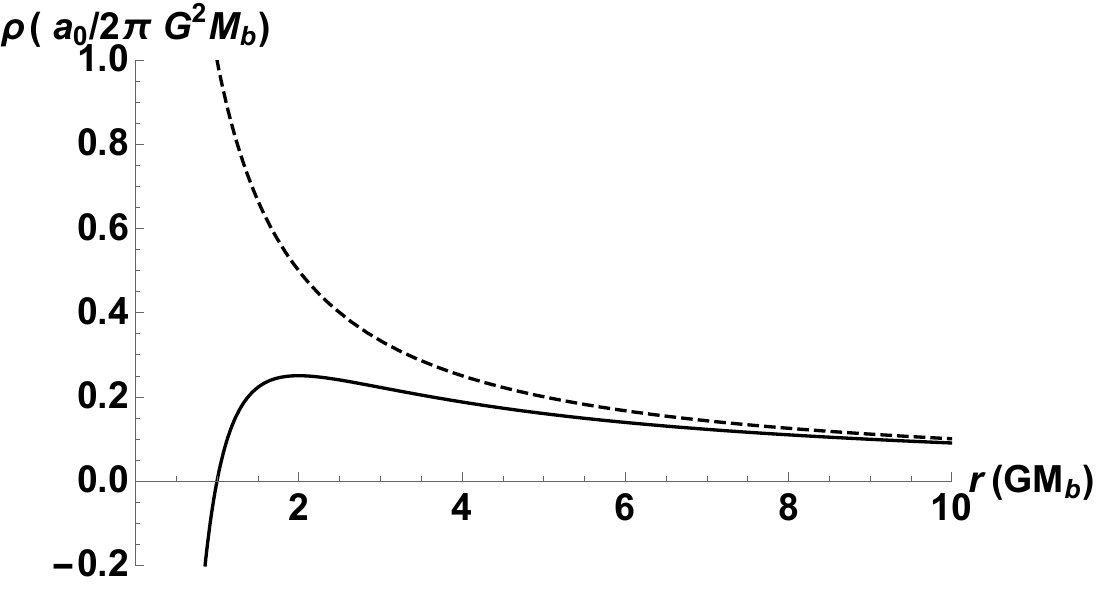}
\end{center}
\caption{\small The mass density of dark matter before and after the inclusion of baryonic matter. The dashed line is the density before the inclusion of baryonic matter  $\rho=a_0/2\pi G r$ which suffers from the core-cusp problem, that is, the density grows up as we approach the centre in contradiction with observations which tell that the density is almost constant in the core. The solid line is the density after the inclusion of baryonic matter, Eq.(\ref{fulldensity}), where one sees that the baryonic matter effect is to break the divergency in the density. What happens to the density inside the black hole horizon is outside the scope of this paper. }\label{fig:core-cusp}
\end{figure}

We found that the inclusion of baryonic matter (the black hole) in our model prevents the cusp to take place in the core. We define the core in our model as the place where the baryonic matter lives, which in our case is the black hole. We include the baryonic matter in our model by returning the black hole to the picture, that is, by returning to our full model with $M_b\neq 0$, where the density reads (see Eq.(\ref{density-P})),
 
  \beq\label{fulldensity}
\rho=\frac{a_0}{2\pi G r}\left(1-\frac{GM_b}{r}\right)
\eeq
which we have plotted in Fig.\ref{fig:core-cusp}.
It can be easily checked from the above equation that the density does not continue growing as we take smaller and smaller $r$'s but reaches a maximum density at the black hole horizon $r=2GM_b$, then it falls off as we move inside the black hole toward the centre till it becomes zero at $r=GM_b$, and then surprisingly for smaller $r$'s it becomes even negative (whatever the interpretation is the inside of the black hole is outside the scope of our paper). 

Although our model differs significantly from real galaxies inside the cores the point we want to highlight in this section is that the divergency in the density (the cusp) disappears as we come close to the region of high concentration of baryonic matter (in our case the black hole) and thus this might give further evidence for the proposal that the core-cusp problem in modelling, or simulating, galaxies will not exist if the baryonic matter is included \cite{deBlok:2009sp,Navarro:1996bv,Pontzen:2011ty}.  

\section{Cutting-off the basic mass density}
\label{cutoff}
The aim of this section is to justify the adjective "basic" in the name  "basic mass density" that we have given to our density function Eq.(\ref{basic density}) and show its exact relation to the NFW profile and other mass profiles in astrophysics and cosmology. In fact, the end of the halo can not be captured by our approach \cite{Haddad:2020koo,Haddad:2020yhe} (that is, we can not find in our approach the mathematical form of the density where it suddenly becomes zero and the dark halo ends) since we are applying the long-wavelength method which, by construction, can not capture sharp changes like this. Therefore, if we want to talk about or find the end of the halo we must resort to introducing cut-offs for our mass density $\rho=a_0/2\pi G r$. There are many ways for introducing cut-offs to mathematical functions; for example, we can use the general form

  \beq\label{}
\rho=\frac{a_0}{2\pi G r\left(1+(r/R)^\alpha\right)^\beta}
\eeq
or  the exponential function,

  \beq\label{}
\rho=\frac{a_0}{2\pi G r}\exp(-r^n/R^n)
\eeq
where the parameter $R$ plays the role of a scale radius which specifies approximately the region where the first behaviour ends and the second begins and where $\alpha$, $\beta$, $n$ are parameters that can be varied so as to obtain the required observational result (see for example the references \cite{Hernquist:1990be,An:2012pv,Zhao:1995cp} and how these parameters were varied to fit the rotation curves of galaxies). Thus, our point is the following: according to our approach  a correct dark matter mass profile must be one of the different ways for choosing a cut-off to the basic density $a_0/2\pi G r$ at large distances with the aim of putting an end to the halo (examples are the NFW, Hernquist \cite{Hernquist:1990be}, some of zhao profiles \cite{An:2012pv,Zhao:1995cp}, and others)\footnote{ Since the basic density $a_0/2\pi G r$ is derived from the Einstein equations under very sensible assumptions we think that a correct mass profile must contain this basic density, or otherwise there must be something wrong with the profile.}.

As for short distances, there the point is different. As stated many times our simplified model is designed to approximate real galaxies only outside the cores. Inside the cores the baryonic mass concentrates and it is expected to influence somehow the distribution of dark matter and change its behaviour compared to its behaviour outside. In this regard, observations show linear rotation curves inside the cores of galaxies  (rigid body behaviour) and hence one concludes that the mass density is constant there (both baryonic and non-baryonic). Thus the full dark matter density can be written  piecewise as,

\beq\label{}
\rho(r) =
 \left\{
     \begin{array}{lr}

                 \rho_0 &  r< r_c  \\
           \frac{a_0}{2\pi Gr\left(1+r/r_s\right)^2}   & r\ge\ r_c \\

     \end{array}
   \right. 
\eeq
where $\rho_0$ is the constant density inside the core and where, as an example, we have cut off the basic density at large distances by the NFW function. The two densities must match at the core radius $r_c$ and so upon requiring this matching (a continuous density) this becomes, 
   \beq\label{}
\rho(r) =
 \left\{
     \begin{array}{lr}

      \frac{a_0}{2\pi Gr_c} +O(r_c/r_s) &  r\le r_c  \\
      \frac{a_0}{2\pi Gr\left(1+r/r_s\right)^2}  & r\ge\ r_c 
        
     \end{array}
   \right.
\eeq
It remains of course to find a physical model inside the core which solves the relevant equations of motion and gives a constant density, a thing which we do not tackle in this work.
 
 One can also, for example, adopt the Burkert profile inside the core, which is almost constant there, upon which the full density reads,
 
 \beq\label{}
\rho(r) =
 \left\{
     \begin{array}{lr}

                 \frac{\rho_0r_c^3}{\left(r+r_c\right)\left(r^2+r_c^2\right)} &  r< r_c  \\ \\
           \frac{a_0}{2\pi Gr\left(1+r/r_s\right)^2}   & r\ge\ r_c \\

     \end{array}
   \right. 
\eeq
where here again we have cut off the basic density by the NFW function at large distances. By requiring continuity at the core radius $r_c$ one obtains,
 \beq\label{}
  \rho(r) =
 \left\{
     \begin{array}{lr}

                 \frac{2a_0r_c^2}{\pi G\left(r+r_c\right)\left(r^2+r_c^2\right)}+O(r_c/r_s)  &  r\le r_c  \\ \\
                 
         \frac{a_0}{2\pi Gr\left(1+r/r_s\right)^2}   & r\ge\ r_c

     \end{array}
   \right. 
\eeq
where one sees that in the full density there are two independent parameters, the core radius $r_c$ (the radius inside which the baryonic mass is enclosed) and the scale radius $r_s$ (roughly speaking, the radius of the halo). Finally, it deserves to be noted that even though the density is continuous at the core radius $r_c$  its derivative is not, a point which should not make us worry since the core surface stands as an interface between two media -- pure dark matter outside the surface and dark matter plus baryonic matter inside -- and hence a discontinuity is expected to appear somehow.

\section {Discussion} 
\label{Diss}

Even though most phenomena concerning galaxies and dark matter  are non-relativistic it is important to put the dark matter in its relativistic framework for the following reasons. First, to see whether dark matter can be indeed fitted within General Relativity, or whether modified gravity theories are needed or not. Second, it is interesting to get an idea of how dark matter behaves in the relativistic regime, whether some new phenomena are expected and how, especially nowadays when particle accelerators are improving and as search for dark matter candidates in particle physics is very active. Third, the dark matter phenomenon must connect at large distances with cosmology (dark energy and the FLRW spacetime) and thus a relativistic framework is no doubt needed. For us, it was the search for a simplified model for galaxies which called for a black hole in the picture (to envelop all the baryonic mass) and the black hole in turn called for the theory of general relativity.

We have seen that the baryonic Tully-Fisher relation \cite{Tully:1977fu,McGaugh:2000sr}, the constant surface density of dark matter  \cite{Kormendy:2004se,Donato:2009ab,K:2020bmx}, the scale-density/scale-radius relation of NFW \cite{Boyarsky:2009rb,Boyarsky:2009af,DelPopolo:2012eb}, all have the same source, namely, the universal constant $a_0$ (see Sec.\ref{review}, Sec.\ref{burk}, and Sec.\ref{NFW} respectively). A question to be asked by right then is: what is the origin of this universal constant? Since acceleration is a result and not a cause the origin must be the mass distribution in space given by the basic density $\rho=a_0/2\pi G r$. Because the dark halo is very large in volume compared to the baryonic volume one can with good reason consider the dark halo as a reservoir and the baryonic volume as the small system (in statistical physics terms), and so the local properties of the reservoir are supposed to be the same at leading order for any reservoir regardless of its size and regardless of the small system; the global properties on the other hand may differ from halo to halo. In this perspective the basic density  $\rho=a_0/2\pi G r$ is a local property of halos (reservoirs) and that is why it is shared by all dark halos. This is the origin of the the universal scaling laws and other relations found in galaxies. Of course, one can try to be more deep and ask about the origin of the basic mass density. Our speculation is that this question lies in particle physics; what is the particle whose gas is pressure-less and can hold together by gravitation to make up a spherical halo with mass density $\rho=a_0/2\pi G r$? It is worth noting in this context that our approach differs from other approaches in that it does not assume anything a priori about the nature of dark matter but puts other physical assumptions (to be mentioned in the next paragraph) as starting points; thus our approach is supposed to be able to predict the nature of dark matter.

The work in this paper provides further evidences for the correctness and importance of the analytical model proposed by the authors in  \cite{Haddad:2020koo,Haddad:2020yhe} for galaxies and galaxy clusters. Many astrophysical and cosmological relations could be derived easily from this model, like the flat rotation curve, the baryonic Tully-Fisher relation, the constant surface density of dark matter, the scale-radius/scale density relation, the rotation curve at intermediate distances (where the baryonic matter is negligible), the core-cusp problem, etc. The model \cite{Haddad:2020koo,Haddad:2020yhe} was derived from very basic and correct physical assumptions and that is why it gave results that are in very good agreement with observations. In the following we list these assumptions. The model assumes that Newtonian gravity is the correct non-relativistic framework and that Einstein equations are the correct relativistic framework. It assumes that the derivative expansion method can be applied validly in galaxies and in galaxy systems and that is seen to be a correct assumption simply by recalling the unit of kilo-parsec used in these systems which equals to $3.1\times 10^{19}$ meter, implying that nothing significantly changes within one kilo-parsec and by recalling the flat rotation curve. It assumes, in accord with observations, that most of the baryonic matter lies inside the cores of galaxies.  It relies on one of the most  well-known results of Gauss's theorem that the gravitational field outside any spherical  symmetric system is the same as the gravitational field of a point particle of the same mass located at the centre.

Finally, we would like to make a comment concerning the reason for using the NFW profile in this paper (and not other profiles) and explaining our general view of things. In fact, we have decided to use the NFW profile because it contains the piece $1/r$ which we claim in this paper to be the basic dark matter density. Our general view of things is as follows: this piece (the $1/r$) is correct at intermediate distances, it fails at short distances since it is not cored, and at large distances it must be cut-off so as to make an end to the halo. In this sense, we are not saying that the NFW is the correct and unique profile; in the paper we have said it clearly that it fails in the core and must be replaced by something else there, like for example, the Burkert profile (see sec.\ref{cutoff}); at large distances the NFW behaviour is a kind of a cut-off to the basic piece ($1/r$) meant to put an end to the halo (see sec.\ref{cutoff}) and in fact there are many other possibilities for making this cut-off other than NFW.

\end{document}